\documentclass[12pt,fullpage]{article}

\usepackage{graphicx}
\usepackage{pgf}
\usepackage{latexsym}
\usepackage{color}
\usepackage{multirow}

\usepackage{amsmath}
\usepackage{amssymb}
\usepackage{amsthm}
\usepackage{amsfonts}
\usepackage{bm}
\usepackage{epstopdf}
\usepackage{kbordermatrix}

\newcommand{\bigCI}{\mathrel{\text{\scalebox{1.07}{$\perp\mkern-10mu\perp$}}}}
\newtheorem{lemma}{Lemma}[section]

\newtheorem{propo}{Proposition}[section]

\newtheorem{teo}[propo]{Theorem}

\begin{document}

\title{\bf Learning binary undirected graph \\ in low dimensional regime}

\author{
{ Daniela De Canditiis}\\	
Istituto per le Applicazioni  del Calcolo "M. Picone",\\
 CNR -  Rome -  Italy\\
{\it d.decanditiis@iac.cnr.it}
}

\date{}

\maketitle

\begin{abstract}
Given a random sample extracted from a Multivariate Bernoulli Variable (MBV), we consider the problem of estimating the structure of the undirected graph for which the distribution is pairwise Markov and the parameters' vector of its exponential form. We propose a simple method that provides a closed form estimator of the parameters' vector and through its support also provides an estimate of the undirected graph associated to the MBV distribution. The estimator is proved to be consistent but it is feasible only in low-dimensional regimes. Synthetic examples illustrates its performance compared with another method that represents the state of the art in literature. Finally, the proposed procedure is used for the analysis of a real data set in the pediatric allergology area showing its practical efficiency.
\end{abstract}

\vspace{2mm}
{\em AMS (2000) Subject Classification}: {Primary: 62G05.  Secondary:  62-07  }

\bigskip


\section{Introduction} \label{sec:intr}
Graphical models are an elegant framework to deal with complex systems of random variables and it is becoming strategic for the statistical analysis of data in a variety of domains such as bioinformatics, image analysis, physics, economics, etc. In many of these contexts one is interested in exploring the complex dependence structure among random variables by using graphical model inference. In   this work we deal with the problem of learning a undirected graph which encodes the conditional dependence relationship between components $(X_1,\ldots,X_p)$ of a Multivariate Bernoulli Variable (MBV).

It is very important to note that the conditional dependence relationship is very different from the marginal dependence relationship and that the former does not imply the second nor vice versa, as pointed out in the well know Yule-Simpson effect \cite{Blyth1972}. More precisely, two variables $ X_i $ and $ X_j $ are conditionally independent (conditioned on the rest of the other system's variables $ X_l $ with $ l \neq i, j $) if their conditional distribution is the product of the conditional marginal distributions, while two variables are independent (in the classical sense, i.e. marginally) if their joint distribution (i.e. the marginal of $ X_i $ and $ X_j $) is the product of the marginals. The concept of conditional independence, being more sophisticated with respect to the marginal one, can capture more fundamental relations between variables and this is the reason why it is becoming central in the analysis of complex system of variables. As an example, consider a data set which consists of recording $p$ simultaneous presence/absence of allergy for $p$ different allergens, it is then possible, to model  the joint distribution of  these $p$ Bernoulli variables as a MBV. Starting from the dataset,  measuring these $p$ Bernoulli variables in $n$ different subjects, one wants to discriminate between direct and indirect association among the different allergens. This is an example of cross-reactivity network between allergens (see \cite{allergology_paper}), where the marginal (indirect) relationship between reaction to different allergens is almost certainly present since the system of $p$ variables is very complex and each variable interacts certainly with the others, and thus we are not interested in it; yet the relationship of conditional (direct) dependence expresses a deeper and more interesting link from the allergological point of view. The statistical task of testing conditional independence
has been extensively studied in various forms within the statistics and econometrics communities for nearly a century, see for example\cite{proceeding2011},  \cite{Canonne2018}, \cite{proceeding2019} and reference therein. However, in this paper we do not propose a new hypothesis test, but in a broader sense we face a parametric estimation problem for an MBV that will have implications on the conditional dependence relationship among components. 

More specifically, MBV admits a parametrization within the framework of exponential families which guarantees a direct interpretation of conditional independence through the exponential family canonical parameters. In this work we are interested not only in the problem of learning the graph underlying the MBV, but we also deal with the problem of learning the parameters' vector of its exponential representation. It will be clear during the exposition that these two problems are strongly connected because the problem of learning the graph is reduced to the problem of learning the support of the parameters' vector of the MBV exponential representation. Hence, we can even say that we're dealing with the problem of learning a factorization of the MBV, which indeed is equivalent to learn the graph structure.

Such a problem has been addressed in the recent statistical literature. For example in \cite{inverse_cov}, given a sample extracted from an MBV, it is proposed to estimate the graph by using the support of the generalized covariance matrix, however this method is applicable only for graph with singleton separator sets (tree being a special case of this class) and moreover this method does not furnish an estimate of the parameters' vector of the exponential MBV representation. On the other hand, all the others existing methodologies  for estimating the parameters' vector make use of a $l_1$- penalized maximum likelihood approach, under sparseness hypothesis on the graph. To be more precise, in \cite{Hofling and Tibshirani 2009}  it is proposed a procedure for solving a class of $l_1$-regularized log likelihood models which estimate the parameters' vector and hence the graph structure of a binary pairwise Markov network. A  binary pairwise Markov network is a MBV with interaction term up to order two. In \cite{Ising_logistic} an $l_1$-regularized logistic regression approach is proposed to learn the signed set of neighbors of each graph's node for an Ising model. An Ising model is a MBV with interaction term up to order two and different support since value 0 is replaced by value -1;  the $l_1$-regularized logistic regression can be slightly modified to obtain an estimate of parameters' vector not only of its signed support, possible modifications are presented in  \cite{Hofling and Tibshirani 2009} as well as in \cite{Io} in the case of symmetric  model (i.e. no first order terms). Moreover it is important to note that all these procedures can be in principle easily extended to general MBV with interaction terms of any order at the price of a severe increase of computational cost; more importantly all these procedures are useful in high dimensional regime, the $l_1$-regularized logistic regression approach of \cite{Ising_logistic} being the most widely used procedure in many different applications.  

Instead of using a maximum likelihood principle, in this paper we propose simple empirical method to estimate the parameters' vector which can work for general MBV. The method is efficient in a low dimensional regime. The great advantage of this procedure lies in its simplicity of calculation, because it provides an estimator in a closed form and hence there is no need for iterative procedures as in the case of maximum likelihood estimators. Moreover, theoretical properties for this estimator are obtained under very general assumption on the underlying MVB, hence there is no need of sophisticated hypothesis  as for the case of maximum likelihood estimators.

The paper is organized as follows. In Section 2 we present population level results, i.e. theoretic properties of a Multivariate Bernulli Variable and set the statistical problem. In Section 3 we review in detail the mechanics of the $l_1$-regularized logistic regression (MLE based) proposed in \cite{Ising_logistic} and we  adapt it to our context. In the same section we introduce our procedure proving a theoretical consistency result. In  Section 4 we present some results on simulated data and finally in Section 5 we apply the proposed procedure to a real case problem.

\section{Mathematical framework}

\subsection{Binary undirected graphs}
For a complete and exhaustive treatment of graphs theory we refer to \cite{Lauritzen}; below we give only definitions and properties necessary for this work.
A finite graph $G=(V,E)$ consists of a finite collection of \textit{nodes} $V=\{1,2...,p\}$  and a collection of \textit{edges} $E \subseteq V\times V$. For the scope of this work, we will consider graphs that are \textit{undirected}, namely graphs whose edges are not ordered, i.e. there is no distinction between the edges $(i,j)$ and $(j,i) \in E$. Moreover, for any $i \in V$ $N(i) := \{j \in V : (i,j) \in E\}$ is the set of neighbours of node $i$. 

In this paper the notion of a graph is used to keep track of the conditional dependence relationship between random variables of a complex system.
By complex system here we mean a jointly distributed  vector of random variables $(X_1,X_2,...,X_p)$ that interact with each other.
%
%
%
 
Associated with an undirected graph $G=(V,E)$ and a system of random variables $X_V$ indexed in the vertexes set $V$ there is a range of different Markov properties which establish how much the graph is explanatory of the conditional independence property of the random variables, see \cite{Lauritzen} for details. Specifically, in this work we deal with systems of random variables which are pairwise Markov with respect to an undirected graph $G=(V,E)$, i.e. it holds
$$
X_i  \bigCI X_j | X_{V\backslash\{i,j\}} \quad \Leftrightarrow \quad (i,j) \notin E,
$$  
which establish conditional independence among two variables $X_i$ and $X_j$ iff their corresponding nodes in the graph $G$ are not connected. Moreover, in \cite{Lauritzen} it is also defined the factorization property of a distribution, specifically a joint distribution factorizes if it can be expressed by an exponential form strictly connected to the graph structure (as the one showed in eq.~(\ref{eq: formula_esponenziale})). The remarkable theorem of Hammersley and Clifford (cfr. Theorem 3.9 in \cite{Lauritzen}) say that for a positive distribution the factorization property is equivalent to the pairwise Markov property. Hence, in this paper, our working hypothesis is that the MBV is positive and it admits an exponential form.  

Our perspective is inferential, therefore, given a statistical sample extracted from the unknown distribution $ f (X_1,\ldots,X_p) $, we are interested into two goals: i) learn the structure of the graph for which the distribution is pairwise Markov ii)  learn the parameters' vector which characterizes its exponential form. In the subsequent section, we explain in detail why these two problems are strongly connected.

\subsection{Multivariate Bernoulli distribution}
In this section we  present some properties of MBV that will  be instrumental for defining the statistical technique discussed in this paper. 
Let $(X_1,\ldots,X_p)$ be a MBV, this means that each variable $X_i$ assumes value in $\{0,1\}$, hence  $(X_1,\ldots,X_p) \in \{0,1\}^p$, the support being of cardinality $2^p$. From a classical point of view, each possible outcome can be identified by the subset $D \subseteq V=\{1,\ldots,p\}$ of variables assuming value 1, with all the others assuming value 0; then the distribution can be expressed by the following formula:
\begin{equation} \label{eq: formula_classica}
p(x_1,\ldots,x_p)= \sum_{D \subseteq V}~ p_D ~ \prod_{i \in D} x_i ~ \prod_{i \notin D} (1-x_i), 
\end{equation}
where $p_D$ is the p.m.f. of configuration $D$, with the constrain $\sum_{D \subseteq V}~ p_D=1$.

For clarity throughout the paragraph we will illustrate the simple case $p=3$, then formula (\ref{eq: formula_classica}) becomes
$$
\begin{array}{ccl}
p(x_1,x_2,x_3)&=&p_{000} (1-x_1)(1-x_2)(1-x_3) + p_{100} x_1(1-x_2)(1-x_3) + p_{010} (1-x_1)x_2(1-x_3) +\\
 & &p_{001} (1-x_1)(1-x_2)x_3
+ p_{110} x_1 x_2(1-x_3) + p_{101} x_1(1-x_2)x_3 +\\
& & p_{011} (1-x_1)x_2 x_3+p_{111} x_1 x_2 x_3, 
\end{array}
$$
where for example $p_{000}=p_D$ with $D=\emptyset$; $p_{011}=p_D$ with $D=\{2,3\}$, ecc...
Very remarkable properties of MBV are discussed and presented in \cite{Wabha2013}; for example: independence and uncorrelatedness are equivalent, both marginal and conditional distributions of subset of variables are still MBV. All these properties resemble that of (MGV) Multivariate Gaussian Variable, making some results that will be drown in the following less surprising.
MBV representation (\ref{eq: formula_classica}), although it is very simple and intuitive, does not offer a direct interpretation of the conditonal dependency among variables. For that reason many authors (see e.g. \cite{Wabha2013} and \cite{Roverato_book}), offer an alternative parametrization of the MBV within the framework of exponential families which guarantees a direct interpretation of conditional independence through the exponential family canonical parameter. More precisely, let us define $\mathcal{P}(V)$ the power set of $V$ and denote $\boldsymbol{\theta}_{\mathcal{P}(V)}=(\theta_D)_{D \subseteq V}$ the parameters' vector which will be used to express  distribution given in (\ref{eq: formula_classica}) in exponential form 

 \begin{equation} \label{eq: formula_esponenziale}
 p(x_1,\ldots,x_p)= exp(\sum_{D \subseteq V} ~ \theta_D ~ \prod_{i \in D} x_i).
 \end{equation}
In the simple case $p=3$, expression (\ref{eq: formula_esponenziale}) becomes
$$
p(x_1,x_2,x_3)= exp \left( \theta_0+\theta_{1} x_1 +\theta_{2} x_2 +\theta_{3} x_3 +\theta_{12} x_1 x_2 +
 \theta_{13} x_1 x_3 +\theta_{23} x_2 x_3 + \theta_{123} x_1 x_2 x_3 \right)
$$
with the obvious notations $\theta_0=\theta_{\emptyset}$, $\theta_{1}=\theta_{\{1\}}$, ecc... Expression  (\ref{eq: formula_esponenziale}), when positive, can be logarithmically transformed to obtain a log-linear model, see for detail \cite{capitolo3} and \cite{Roverato_book}. 
Since equations (\ref{eq: formula_classica}) and (\ref{eq: formula_esponenziale}) are equivalent there is a one-to-one relationship  between the probability vector $\boldsymbol{\pi}_{\mathcal{P}(V)}=(p_D)_{D \subseteq V}$ used in (\ref{eq: formula_classica}) and the parameters' vector $\boldsymbol{\theta}_{\mathcal{P}(V)}=(\theta_D)_{D \subseteq V}$ used in (\ref{eq: formula_esponenziale}). In the following we make this relation explicit.
Let us first define the \emph{ zeta} matrix, $\mathcal{Z}_i$, and its inverse $\mathcal{M}_i$, called the \emph{ M\"obius }matrix associated to the set $\{ i\}$:
\begin{equation}
\mathcal{Z}_i= \kbordermatrix{&\emptyset &\{i\}\cr
                \emptyset &1 &  1 \cr
                \{i\}     &0 & 1 }  \quad \quad 
\mathcal{M}_i=\kbordermatrix{&\emptyset &\{i\}\cr
                \emptyset &1 &  -1 \cr
                \{i\}     &0 & 1 }.
\end{equation}  
Let us now define the \emph{ zeta } and \emph{M\"obius} matrices associated to $V=\{1,\ldots,p\}$: 
\begin{equation}
\mathcal{Z}= \otimes_{i \in V} \mathcal{Z}_i
                \quad \quad 
\mathcal{M}=\otimes_{i \in V} \mathcal{M}_i.
\end{equation}
In the case $p=3$, hence $V=\{1,2,3\}$, the \emph{zeta} matrix is:
$$
\mathcal{Z}_V= \kbordermatrix{&\emptyset &\{1\}&\{2\}&\{3\}&\{12\}&\{13\}&\{23\}&V \cr
                \emptyset  &1 & 1 & 1 & 1 & 1 & 1 & 1 & 1 \cr
                \{1\}      &0 & 1 & 0 & 0 & 1 & 1 & 0 & 1 \cr
                \{2\}      &0 & 0 & 1 & 0 & 1 & 0 & 1 & 1 \cr
                \{3\}      &0 & 0 & 0 & 1 & 0 & 1 & 1 & 1 \cr
                \{12\}     &0 & 0 & 0 & 0 & 1 & 0 & 0 & 1 \cr
                \{13\}     &0 & 0 & 0 & 0 & 0 & 1 & 0 & 1 \cr
                \{23\}     &0 & 0 & 0 & 0 & 0 & 0 & 1 & 1 \cr 
                V          &0 & 0 & 0 & 0 & 0 & 0 & 0 & 1 \cr
                },
$$

and its inverse is:
$$
\mathcal{M}_V= \kbordermatrix{&\emptyset &\{1\}&\{2\}&\{3\}&\{12\}&\{13\}&\{23\}&V \cr
                \emptyset  &1 & -1 & -1 & -1 & 1 & 1 & 1 & -1 \cr
                \{1\}      &0 & 1 & 0 & 0 & -1 & -1 & 0 & 1 \cr
                \{2\}      &0 & 0 & 1 & 0 & -1 & 0 & -1 & 1 \cr
                \{3\}      &0 & 0 & 0 & 1 & 0 & -1 & -1 & 1 \cr
                \{12\}     &0 & 0 & 0 & 0 & 1 & 0 & 0 & -1 \cr
                \{13\}     &0 & 0 & 0 & 0 & 0 & 1 & 0 & -1 \cr
                \{23\}     &0 & 0 & 0 & 0 & 0 & 0 & 1 & -1 \cr 
                V          &0 & 0 & 0 & 0 & 0 & 0 & 0 & 1 \cr
                }.
$$
{\bf \emph{Remark 1.}}   It is worthwhile to observe that in each row of $\emph{zeta}$ matrix there is 0 or 1 if the set corresponding to that row is a subset of the set corresponding to the column; for example, first line has 1 in each position because $\emptyset$ is a subset of every $D \subseteq V$, while fourth line has 1 only in the positions corresponding to the sets $\{3\}$, $\{13\}$, $\{23\}$ and  $V=\{123\}$ which contains  $\{3\}$, ecc... While in \emph{ M\"obius} matrix in each column there is 0 if the set corresponding to that column is not a subset of the set corresponding to the row and there is $1$ or $-1$ if it is a subset, the sign alternating between sets of odd and even cardinality.

We can now state the following lemma:

\begin{lemma} \label{lemma_theta_pi}
Let $\boldsymbol{\pi}_{\mathcal{P}(V)}$ be the positive probability vector of the MBV expressed in eq.(\ref{eq: formula_classica}) and let  $\boldsymbol{\theta}_{\mathcal{P}(V)}$ be the parameter vector of the same MBV expressed in the exponential form of eq. (\ref{eq: formula_esponenziale}), then it holds that:
$$
\boldsymbol{\pi}_{\mathcal{P}(V)}=exp(\mathcal{Z}^t_V \boldsymbol{\theta}_{\mathcal{P}(V)}) \quad \quad \mbox{and} \quad \quad \boldsymbol{\theta}_{\mathcal{P}(V)}=\mathcal{M}^t_V log(\boldsymbol{\pi}_{\mathcal{P}(V)}),
$$
where $log$ and $exp$ are taken entrywise and $*^t$ is the transpose of $*$.. 
\end{lemma}
The proof of this Lemma is in \cite{Roverato_book} Par. 4.3.

Note that $\boldsymbol{\pi}_{\mathcal{P}(V)}$ is component-wise positive, i.e. its component are probabilities, $p_D >0$, while  the vector $\boldsymbol{\theta}_{\mathcal{P}(V)}$ has real value components.

Again in the case $p=3$ we use Lemma \ref{lemma_theta_pi} to explicitly give the relation between the two  parameters' vectors:
$$
\left( \begin{array}{c} \theta_{0} \\ \theta_{1} \\\theta_{2} \\\theta_{3} \\ \theta_{12} \\ \theta_{13} \\ \theta_{23} \\
\theta_{123} \\ \end{array} \right) = 
\left[ \begin{array}{cccccccc}
 1 & 0 & 0 & 0 & 0 & 0 & 0 & 0 \\ 
 -1 & 1 & 0 & 0 & 0 & 0 & 0 & 0 \\
-1 & 0 & 1 & 0 & 0 & 0 & 0 & 0\\
-1 & 0 & 0 & 1 & 0 & 0 & 0 & 0\\
1 & -1 & -1 & 0 & 1 & 0 & 0 & 0\\
1 & -1 & 0 & -1 & 0 & 1 & 0 & 0\\
1 & 0 & -1 & -1 & 0 & 0 & 1 & 0\\
-1 & 1 & 1 & 1 & -1 & -1 & -1 & 1
\end{array} \right] \left( \begin{array}{c} log( p_{000}) \\ log( p_{100}) \\log( p_{010}) \\log( p_{001}) \\ log( p_{110}) \\ log( p_{101}) \\ log( p_{011}) \\
log( p_{111}) \\ \end{array} \right)
$$
It is interesting, to express the above equality component-wise:
\begin{equation} \label{eq:caso3}
\left\{ \begin{array}{ccl}
\theta_0 & = &log(p_{000})       \\
\theta_{1} & = &  log(p_{100}/ p_{000})    \\
\theta_{2} & = &   log(p_{010}/ p_{000})      \\
\theta_{3} & = &   log(p_{001} / p_{000})      \\
\theta_{12} & = &   log(p_{110} p_{000} / p_{100} p_{010})      \\
\theta_{13} & = &  log(p_{101} p_{000} / p_{100} p_{001})      \\
\theta_{23} & = &  log(p_{011} p_{000} / p_{010} p_{001})      \\
\theta_{123} & = &  log(p_{111} p_{100} p_{010} p_{001} / p_{000} p_{110} p_{101} p_{011}).  \\
\end{array}   \right.
\end{equation}

We can now state the main result which formally establishes the connection between conditional independence relationship (hence graph structure) and the support of vector $\boldsymbol{\theta}_{\mathcal{P}(V)}$.

\begin{teo} \label{teo:Theorem4.2_roverato}
For a MBV $(X_1,\ldots,X_p)$ with distribution given in eq.(\ref{eq: formula_classica}) with positive probability $\boldsymbol{\pi}$, let $\boldsymbol{\theta}=  \mathcal{M}^t_V log(\boldsymbol{\pi})$. Then, for a pair of disjoint non-empty subsets $A$ and $B$ of $V$ the following conditions are equivalent
\begin{itemize}
\item[(i)] $X_A \bigCI X_B | X_{V \setminus (A \cup B)}$

\item[(ii)] for every $D \subseteq V$ such that both $A \cap D \neq \emptyset$ and  $B \cap D \neq \emptyset$ it holds that $\theta_D=0$
\end{itemize}
\end{teo} 

This theorem coincides with Theorem 4.2 of \cite{Roverato_book} from where its proof can be taken. For clarity, here we only give an idea of the proof for the simple case $p=3$, proving that $X_1 \bigCI X_2 | X_3$ iff $\theta_{12}=0 \land \theta_{123}=0$.
From equation~(\ref{eq:caso3}) we have that
$$
\theta_{12}= log \left( \frac{p_{110} p_{000} }{p_{100} p_{010}} \right)=log \left( \frac{p_{11|0} p_{00|0}}{p_{10|0} p_{01|0}} \right),
$$
where, for example, $p_{11|0}$ is a short for $P(X_1=1,X_2=1|X_3=0)$, hence 
\begin{equation} \label{eq:oddratio}
\theta_{12}= 0 \quad \leftrightarrow \quad  \frac{P(X_1=1,X_2=1|X_3=0) P(X_1=0,X_2=0|X_3=0)}{P(X_1=1,X_2=0|X_3=0) P(X_1=0,X_2=1|X_3=0)} =1,
\end{equation}
expression (\ref{eq:oddratio}) is the conditioned \emph{odd ratio} of variables $X_1$ and $X_2$. Since it is well known that odd ratio equals to one is equivalent to independence, we have that  
$$
 \theta_{12}= 0  \quad \leftrightarrow \quad X_1 \bigCI X_2 |X_3=0.
$$
With condition $ \theta_{12}= 0 $ true, we can rewrite the last equation of (\ref{eq:caso3}) obtaining
$$
\theta_{123}= log \left(\frac{p_{111} p_{001}  }{ p_{101} p_{011}} \right)- \underbrace{log \left( \frac{p_{110} p_{000}}{p_{100} p_{010}} \right)}_{\theta_{12}=0} =
log \left( \frac{p_{11|1} p_{00|1}}{p_{10|1} p_{01|1}} \right),
$$
hence with $\theta_{12}= 0 $, with the same argument we get $ \theta_{123}= 0 \quad \leftrightarrow \quad X_1 \bigCI X_2 |X_3=1$.

Theorem \ref{teo:Theorem4.2_roverato} is strategic for the problem of learning conditional independence relationship among variables of a MBV, indeed if we consider $A=\{i\}$ and $B=\{j\}$, it follows that $X_i \bigCI X_j | X_{V \setminus \{i,j\}}$ iff $\theta_D=0$ for all $D$ super set of $\{i,j\}$.
Since for most application, it is not meaningful to include higher-order interaction terms without incorporating the lower-order interactions, here and in the majority of literature, the hierarchical hypothesis is considered. 
In particular, for a  hierarchical exponential model, whenever an interaction term is fixed to zero then all higer-order iteraction terms involving the same variables are also zero. More formally, for a hierarchical exponential model, for every nonempty $D \subseteq V$  it holds:

\begin{equation} \label{eq:group_spar}
\theta_D=0 ~~\rightarrow ~~\theta_E =0 ~\mbox{for all} ~E \supseteq D
\end{equation}

or equivalently,
$$
\theta_D \neq 0~~ \rightarrow~~ \theta_E \neq 0 ~\mbox{for all}~ E \subseteq D~ \mbox{with}~ E \neq \emptyset.
$$
With this hypothesis, in the simple case $p=3$,
the model 
$$
p(x_1,x_2,x_3)= exp \left( \theta_0+\theta_{1} x_1 +\theta_{2} x_2 +\theta_{3} x_3  +
 \theta_{13} x_1 x_3 +\theta_{23} x_2 x_3 \right)
$$
is plausible, while the following one is not
$$
p(x_1,x_2,x_3)= exp \left( \theta_0+\theta_{1} x_1 +\theta_{2} x_2 +\theta_{3} x_3  +
 \theta_{13} x_1 x_3 +\theta_{23} x_2 x_3 +\theta_{123} x_1 x_2 x_3 \right).
$$

\section{Method for learning a binary undirected graph}
There are many proposals in statistical literature for learning binary graphs, but in this work we are interested only in those considering the problem of learning graph structure as well as  estimating parameter $\boldsymbol{\theta}_{\mathcal{P}(V)}$. Of course, the second task is more ambitious because, once we have a good estimator $\boldsymbol{\hat{\theta}}_{\mathcal{P}(V)}$,  by applying  Theorem~\ref{teo:Theorem4.2_roverato}, the graph structure is obtained considering its support, i.e.  $\hat{E}=\{ (i,j): \hat{\theta}_{ij}  \neq 0\}$. Hence we formulate our problem as that of estimating parameter $\boldsymbol{\theta}_{\mathcal{P}(V)}$ of a hierarchical and positive MBV, given a sample of $n$ independent realizations, $\{(X_1^{(i)},...,X_p^{(i)})\}_{i=1,...,n}$ .

\subsection{Logistic regression approach} \label{sec:logistic}

One of the most widely used method is the one proposed in \cite{Ising_logistic} where an $l_1$-regularized logistic regression is applied for learning an Ising model under high dimensional regime. The Ising model is a MBV with no-interaction terms of order greater than two and with support $\{-1,+1\}^p$ instead of $\{0,1\}^p$. 
Here we briefly describe their proposal to adapt it to our context. In \cite{Ising_logistic} the authors consider the conditional distribution of variable $X_j$ given the rest $X_{V\setminus \{j\}}$. From the properties of MBV it is true that $p(X_j | X_{V\setminus \{j\}})$ is still a Bernoulli variable with probability of success given by the following expression
\begin{equation} \label{eq:conditioned}
p(X_j=1| X_{V\setminus \{j\}})=\frac{exp(\sum_{D \ni j } ~ \theta_D ~ \prod_{i \in D \setminus \{j\}} x_i)}{1+exp(\sum_{D \ni j} ~ \theta_D ~ \prod_{i \in D \setminus \{j\}} x_i)}.
\end{equation}

Then, it is possible to see the variable $X_j$ as the response  in a logistic regression problem in which all the other variables $X_{V\setminus \{j\}}$ as well as all their possible interaction terms play the role of covariates.
Under this set-up, the method for estimating the neighborhoods of node $j$ is based on computing a grouped  Lasso-regularized logistic regression with group sparsity governed by condition in eq.~(\ref{eq:group_spar}).

In the simple case $p=3$, for $j=1$ equation (\ref{eq:conditioned}) reduces to
\begin{equation}
p(X_1=1| x_2,x_3)=\frac{exp \left(\theta_1 + \theta_{12} x_2+ \theta_{13}x_3+\theta_{123} x_2 x_3 \right)}{1+exp \left(\theta_1 + \theta_{12} x_2+ \theta_{13}x_3+\theta_{123} x_2 x_3 \right)},
\end{equation}
and the group structure in this case is $\{\theta_{12}, \theta_{123}\} \cup \{\theta_{13}, \theta_{123}\} =\{ \theta_{12}, \theta_{13}, \theta_{123} \}$.

Of course to learn all the graph it is necessary to perform such analysis for each node $j \in V$. 
However, since in each logistic regression, the number of covariates is $ 2^{p-1} $ this method can become expensive from a computational point of view and this is why in the literature it has been explored only for Ising model, i.e. models without interactions of order higher than two; for such a model  the representation given in eq.~(\ref{eq: formula_esponenziale}) simplifies to
$$
p(x_1,\ldots,x_p)= exp(\theta_0+ \sum_{i \in V} \theta_i x_i + \sum_{i <j}  \theta_{ij} ~  x_i x_j).
$$ 
and the logistic regression for the generic variable $X_j$ simplifies to
$$ 
p(X_j=1| X_{V\setminus \{j\}})=\frac{exp(\theta_j + \sum_{i \neq j} \theta_{ij} x_i)}{1+exp(\theta_j + \sum_{i \neq j} \theta_{ij} x_i)}.
$$
 
Under a sparsity hypothesis on the graph structure, the following penalized maximum likelihood  estimator is evaluated for each node $j \in V$
\begin{equation}\label{eq:unalogistic}
\hat{\theta}_{\cdot j}=argmin_{\theta}  \left\{ -\frac{1}{n} \sum_{k=1}^n  log(p(x_j^{(k)}| x_1^{(k)}, x_{j-1}^{(k)}, x_{j+1}^{(k)},...,x_p^{(k)})+\lambda \sum_{i \neq j} |\theta_{ij}| \right\}
\end{equation}
 
As a theoretical support of such method,  in \cite{Ising_logistic}, under certain assumptions, the authors prove that solution  of (\ref{eq:unalogistic}) consistently estimates ${N}_{\pm}(j)=\{ sign( {\theta}_{ij}): i \in N(j)\}$ the signed set of node $j$ neighbourhood. The reason why the authors do not consider $\hat{\theta}_{\cdot j}$ as estimate of the  $\theta_{\cdot j}$ lies in the fact that they solve problem (\ref{eq:unalogistic}) for each node $ j \in V $ independently of the other nodes so that $ \hat{\theta}_{ij} \neq \hat{\theta}_{ji}$.
 For this reason in  \cite{Hofling and Tibshirani 2009} two procedures for symmetrizing this method are proposed. 
The first procedure works in the following way:
\begin{equation} \label{eq: N-L-m}
\hat{\theta}_{ij}= \hat{\theta}_{ji}=\left\{ \begin{array}{cl}
\hat{\theta}_{ij} & if |\hat{\theta}_{ij}| > |\hat{\theta}_{ij}| \\
\hat{\theta}_{ji} & if |\hat{\theta}_{ij}| \leq |\hat{\theta}_{ij}|
\end{array}   \right. ,
\end{equation}
similarly, the second procedure works in the following way:
\begin{equation} \label{eq: L-N-M}
\hat{\theta}_{ij}=\hat{\theta}_{ji}= \left\{ \begin{array}{cl}
\hat{\theta}_{ij} & if |\hat{\theta}_{ij}| < |\hat{\theta}_{ij}| \\
\hat{\theta}_{ji} & if |\hat{\theta}_{ij}| \geq |\hat{\theta}_{ij}|
\end{array}   \right. .
\end{equation}
In \cite{Hofling and Tibshirani 2009} these procedures are referred "Wainwright-min" and "Wainwright-max", and the second is proved to be always superior to the first. That's why in this paper we consider only procedure (\ref{eq: L-N-M}) but we call it L-N-M (Logistic-Neighborhood-Max). This is the procedure used in Section \ref{sec:numeric} for comparisons.


\subsection{The proposed method} \label{sec:metodo_parametri}
In this section we propose a new procedure which has the advantage to be simple and therefore computationally much more convenient with respect to the N-L-M method, moreover its theoretical property are obtained under much more general conditions. 
Before presenting our procedure, let us state the following lemma at the population level

\begin{lemma} \label{lemma: population}
Let $\boldsymbol{\pi}_{\mathcal{P}(V)}$ be the positive probability vector of MBV expressed in eq.(\ref{eq: formula_classica}) and let  $\boldsymbol{\theta}_{\mathcal{P}(V)}$ be the parameter vector of its exponential form expressed in eq. (\ref{eq: formula_esponenziale}), then for each $D \in \mathcal{P}(V)$ it holds

\vspace{0.5cm}

\emph{i)}  $p^D_{min}=min_{D' \subseteq D} ~ p_{D'} > 0 \quad   \mbox{and} \quad  p^D_{max}=max_{D' \subseteq D} ~ p_{D'} <1$

\vspace{0.5cm}

\emph{ii)}  $\theta_D=\sum_{i=1}^{2^{|D|-1}} log \frac{p_{D_{e,i}}}{p_{D_{o,i}}}$ 

where  $D_{e} \subseteq D$ s.t. $|D \setminus D_e|$ is even and $D_{o} \subseteq D$ s.t. $|D \setminus D_o|$ is odd ($|\star|$ cardinality of set $\star$).
\end{lemma}
 
 \emph{Proof.} Since $\boldsymbol{\pi}_{\mathcal{P}(V)}$ is positive by hypothesis we have $0<p_D <1$ ,$\forall D \subseteq V$, hence \emph{i)} is easily proved by contradiction.
 To prove the second claim, we use the \emph{ M\"obius } inversion formula of Lemma~\ref{lemma_theta_pi}, i.e. $\boldsymbol{\theta}_{\mathcal{P}(V)}=\mathcal{M}^t_V log(\boldsymbol{\pi}_{\mathcal{P}(V)})$, and Remark 1 hence it holds 
$$
\theta_{D}=\sum_{D' \subset D} (-1)^{|D \setminus D'|} log(p_{D'}).
$$
Since the number of subsets $D'$ of $D$ such that $|D \setminus D'|$ is even is equal to the number of subsets $D'$ of $D$ such that $|D \setminus D'|$ is odd, by using $log$ function properties we easily obtain claim \emph{ii)}. $\blacksquare$

Let us come back to the inferential problem. Given a sample of size $n >p$, let us define, for each $D \in  \mathcal{P}(V)$, the empirical frequency by the following formula
\begin{equation} \label{eq: emp_freq}
\hat{p}_D^{(n)}=\frac{1}{n} \sum_{i=1}^n \mathbf{1}(X_j^{(i)}=1, \forall j \in D, X_j^{(i)}=0, \forall j \notin D ),
\end{equation}
 where $\mathbf{1}(\cdot) $ is the indicator function. We can now explicitly give expression of the proposed estimator, 
\begin{equation} \label{eq:est_theta}
\hat{\theta}_D^{(n)}=\sum_{i=1}^{2^{|D|-1}} log \left( \frac{\hat{p}^{(n)}_{D_{e,i}}}{\hat{p}^{(n)}_{D_{o,i}}} \right)
\end{equation} 
with $\hat{p}^{(n)}_D$ given in (\ref{eq: emp_freq}).

Using the empirical formulation above and the low of large numbers, we can now state a consistency result for the proposed estimator.

\begin{teo}\label{teo:consistenza}
Given a sample of size $n>p$ of a positive MBV expressed in exponential form by eq.(\ref{eq: formula_esponenziale}), $\forall D \in \mathcal{P}(V)$ it holds

\vspace{0.5cm}

\emph{i)} $lim_{n \rightarrow \infty} \hat{\theta}_D^{(n)}=\theta_D, \mathbb{P}-a.s.$

\vspace{0.5cm}

\emph{ii)} $lim_{n \rightarrow \infty} \mathbb{E}[(\hat{\theta}_D^{(n)})^k]=\theta_D^k, \mbox{for any}~ k \geq 1$. In particular, $lim_{n \rightarrow \infty} \mathbb{V}ar[(\hat{\theta}_D^{(n)})]=0$
\end{teo}

 \emph{Proof.}  By the strong law of the large numbers, as $n \rightarrow \infty$, it follows $\hat{p}^{(n)}_{D_{e,i}} \rightarrow p_{D_{e,i}}$ and $\hat{p}^{(n)}_{D_{o,i}} \rightarrow p_{D_{o,i}}$, $\mathbb{P}-a.s.$. We then prove \emph{i)} by the continuity of the logarithm. 
In order to prove the second claim, let us first give the following elementary inequalities:
 $$
 \frac{p^D_{min}}{p^D_{max}} \leq \frac{\hat{p}^{(n)}_{D_{e,i}}}{\hat{p}^{(n)}_{D_{o,i}}} \leq \frac{p^D_{max}}{p^D_{min}},
 $$
 which holds $\mathbb{P}-a.s.$ for any $n >p$. Since the logarithm is increasing, it then follows
 $$
 -2^{|D|-1} log \frac{p^D_{min}}{p^D_{max}} \leq \hat{\theta}^{(n)} \leq 2^{|D|-1} log \frac{p^D_{max}}{p^D_{min}}, ~ \mathbb{P}-a.s.  ~
 $$
 i.e.
 $$
 |\hat{\theta}^{(n)}| \leq 2^{|D|-1} log \frac{p^D_{max}}{p^D_{min}},  ~ \mathbb{P}-a.s. ~ \mbox{for any } ~ n>p.
 $$
 Claim \emph{ii)} then follows by \emph{i)} and the dominated convergence theorem. $\blacksquare$

The proposed method can be summarized into three steps, the last one being necessary only in the case of sparseness hypothesis on the graph structure:
 
\begin{enumerate}
\item[step 1:] evaluate $\hat{\boldsymbol{\pi}}_{\mathcal{P}(V)}=(\hat{p}_D)_{D \subseteq V}$ by (\ref{eq: emp_freq})
\item[step 2:] evaluate $\hat{\boldsymbol{\theta}}=\mathcal{M}^t_V log(\hat{\boldsymbol{\pi}})$
\item[step 3:] perform a threshold on entries of $\hat{\boldsymbol{\theta}}$  
\end{enumerate}

This learning procedure is simple and its computational cost is really negligible with respect to the iterative method one has to adopt to solve problem in eq. (\ref{eq:unalogistic}); on the other hand, the proposed procedure can be applied only in low dimensional regime, i.e. when the number of data is much higher than the dimension of the problem ($n >>p$). This last limitation is due to the fact that small errors of approximation of the empirical frequencies as estimates of the true frequencies become large approximation errors in the estimate of $\boldsymbol{\theta}$, due to the logarithm's derivative.
Another advantage of the proposed procedure with respect to the N-L-M method is that one can incorporate into estimator $\hat{\boldsymbol{\theta}}_{\mathcal{P}(V)}$ any a prior knowledge of the true $\boldsymbol{\theta}_{\mathcal{P}(V)}$; for example if one knows that terms above some degree are zero, one can set them to zeros; if one know that the maximum degree of $j$-th graph node is $d$, then in the third step of the procedure one can set to zero the entries of vector $(\hat{\theta}_{ij})_{i \neq j}$ which are below the $p-d/p$-th empirical quantile.
When a priori information are not sufficient for the choice of threshold in the third step, it is necessary to perform a model selection procedure, as CV for example,   being this always true for choice of the regularization parameter $\lambda$ in  N-L-M procedure.

Finally,  since we have stressed the analogies at the population level between MBV and MGV, it is also worthwhile to stress that the proposed method is analogous to the method for learning Gaussian Graphical model, which consists of the following three steps: 

\begin{itemize}
\item[step 1:] evaluate the empirical covariance matrix
\item[step 2:] numerically invert the empirical covariance matrix to get an estimate of the precision matrix
\item[step 3:] perform a threshold on the precision matrix elements.
\end{itemize}

This method is described in detail in subsection 7.3.2 of \cite{Giraud}, however the analogy with the proposed one is very clear.

\section{Numerical experiments} \label{sec:numeric}

In this section we show some numerical experiments to study the performance of the proposed method. 
Before presenting results it is necessary to specify indexes we used to measure performance.

Since we are interested both in reconstructing the structure of the graph and in estimating the parameters' vector, we calculate two different indexes of performance.
The first index measures how the method correctly estimates the structure of the graph and it is defined as:
\begin{equation}  \label{eq:accuracy}
accuracy= (TP + TN) / (TP + TN + FN + FP),
\end{equation}
where \begin{itemize}
\item[]$TP$ is the number of edges present in the graph and correctly identified (i.e. $\theta_{ij} \neq 0 \wedge \hat{\theta}_{ij} \neq 0$),
\item[]$TN$ is the number of edges not present in the graph and correctly identified  (i.e. $\theta_{ij} = 0 \wedge  \hat{\theta}_{ij}=0$),
\item[]$FN$ is the number of edges present in the graph and not correctly identified (i.e. $\theta_{ij} \neq 0 \wedge  \hat{\theta}_{ij} =0$) and
\item[]$FP$ is the number of edges not present in the graph and not correctly identified  (i.e. $\theta_{ij} =0\wedge  \hat{\theta}_{ij} \neq 0$).
\end{itemize}
Note that measure in (\ref{eq:accuracy}) is a scaled measure inherit from the binary classification literature, $0 \leq accuracy \leq 1$, being more accurate methods with higher accuracy.

The second index measures how the method correctly estimates the parameters' vector and it is defined as the relative $l_2$-norm error:
\begin{equation}
Err= \|\boldsymbol{\theta}-\hat{\boldsymbol{\theta}} \|_2 / \| \boldsymbol{\theta} \|_2.
\end{equation}

Let us describe the specific setting we chose for numerical experiments. For computational reasons, being the L-N-M method too heavy for a general MBV, we concentrate our attention on model with only  second order interactions.  We propose three examples of different sizes, namely $p=5,10,15$. For the first case $p=5$, we considered example proposed in \cite{Ising_logistic} where 6 out of 10 parameters $\theta_{ij}$ are randomly chosen with \emph{mixed coupling}, i.e. $\theta_{ij}=\pm 0.5$ with equal probability. The second and the third examples, are obtained analogously but with different degree of sparsity. Specifically, in the second example 12 out of 45 parameters $\theta_{ij}$ are no zero, while in the third example 18 out of 105.  For each of the examples we consider five different sample sizes all respecting a low dimensional regime.
For the proposed procedure the threshold in step 3 was chosen among few quantiles (0.2, 0.4, 0.5, 0.6, 0.7) using the a prior sparseness hypothesis, since this information does not offer a way to chose the regularization parameter $\lambda$ for the N-L-M method, for the latter we applied a 10-fold CV procedure to select the best $\lambda$ for each node.

Results are reported in Table \ref{tab: result}, along with the run time of both methods on a workstation i7 8700. It is clear that the proposed procedure does not improve in terms of accuracy, but it furnishes very important improvements both in terms of estimation error and computational time, more significant the higher the sample size is. From our experimentations it comes out that the proposed  procedure become competitive in low dimensional regime when $n > 30 p$, when this requirement is not full fished the procedure is not competitive (results not showed).
\begin{table}
\caption{mean accuracy, error and runtime(sec) over 10 independent simulations. Results are obtained using p=5 and different sample sizes. }
\label{tab} \label{tab: result}
\resizebox{12cm}{!}{
\begin{tabular}{c|cc|cc|cc} 
\hline
 n & \multicolumn{2}{c}{accuracy} & \multicolumn{2}{c}{Err} & \multicolumn{2}{c}{runtime(sec)} \\ 
\hline\noalign{\smallskip}
&  M-I & L-N-M  &   M-I & L-N-M  &  M-I & L-N-M  \\ 
\hline
p=5 & & & & & & \\
 n=150   & 0.86(0.03) &     0.84(0.04) &     0.95(0.56)  &   0.99(0.01) &   0.00(0.01)  &   13.57(0.47) \\
  n=300   &0.88(0.03) &     0.89(0.04) &     0.42(0.37)  &   0.98(0.01) &   0.00(0.00)  &   4.74(0.14) \\
  n=500   &0.92(0.03) &     0.95(0.03) &     0.25(0.05)  &   0.97(0.01) &   0.00(0.00)  &   5.09(0.16) \\
 n=1000   & 0.93(0.02) &     0.99(0.02) &     0.18(0.07)  &   0.96(0.00) &   0.00(0.00)  &   5.88(0.16) \\
 n=5000   & 0.94(0.01) &     1.00(0.01) &     0.09(0.01)  &   0.95(0.00) &   0.00(0.00)  &   13.58(0.47) \\
 
\hline
p=10 & & & & & & \\
n= 300     & 0.99(0.00) &     0.99(0.00) &     0.78(0.28)  &   0.99(0.00) &   0.01(0.01)  &   342.51(5.16) \\
n= 600     &  0.99(0.00) &     1.00(0.00) &     0.34(0.20)  &   0.98(0.00) &   0.02(0.00)  &   12.34(0.24) \\
n= 900     &  0.99(0.00) &     1.00(0.00) &     0.38(0.22)  &   0.98(0.00) &   0.02(0.00)  &   13.57(0.32) \\
n= 10000   &  0.99(0.00) &     1.00(0.00) &     0.34(0.05)  &   0.97(0.00) &   0.11(0.00)  &   58.74(0.81) \\
n=50000    &  1.00(0.00) &     1.00(0.00) &     0.12(0.03)  &   0.97(0.00) &   0.35(0.04)  &   342.51(5.16) \\
 \hline
 p=15 & & & & & & \\
n=500    & 1.00(0.00) &     1.00(0.00) &     0.39(0.14)  &   0.99(0.00) &   0.85(0.14)  &   1187.19(18.74) \\
n=1000   & 1.00(0.00) &     1.00(0.00) &     0.37(0.11)  &   0.99(0.00) &   1.07(0.02)  &   19.86(0.24) \\
n=10000  & 1.00(0.00) &     1.00(0.00) &     0.23(0.02)  &   0.98(0.00) &   5.93(0.03)  &   112.79(1.43) \\
n=50000  & 1.00(0.00) &     1.00(0.00) &     0.20(0.00)  &   0.98(0.00) &   13.90(0.15)  &   613.39(8.38) \\
n=100000 & 1.00(0.00) &     1.00(0.00) &     0.20(0.00)  &   0.98(0.00) &   35.95(0.96)  &   1187.19(18.74)\\
\end{tabular}
}
\end{table}

Matlab codes to reproduce results are available at http://www.iac.cnr.it/~danielad/software.html.

\section{Real data application}
\begin{figure} \label{fig:real_data}
\centering{
\includegraphics[width=0.75\columnwidth]{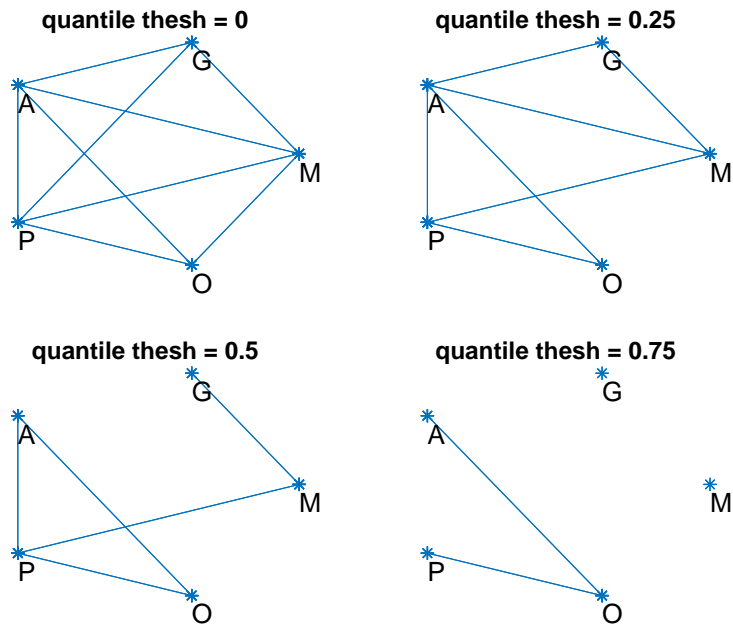}}
\caption{ Aeroallergens network estimated from pediatric records of 200 children with symptoms of respiratory track inflammation. Allergens are abbreviated like grasses (G), dust mites (M), olea (O), parietaria (P) and  alternaria (A). Four different solutions are obtained applying our procedure with four different thresholds (expressed in quantile of estimated $\hat{\theta}_{ij}$)}
\end{figure}

Allergy is the result of an inadequate immune response with a genetic or atopy predisposition, at least 20\% of the population of industrialized countries suffers from different forms of allergies. The development of allergy is a complicated and not completely understood process, a step towards  understanding it is offered by the molecular analysis of allergens. However, molecular analysis requires time and economic resources, so before proceeding with this type of investigation researcher try to understand through cross-reactivity studies which associations exist between different allergens. The association between one allergen and another can be interpreted as a relationship of conditional dependence between the variables that record presence/absence of allergies for different allergens. Therefore, the data analysis presented in this section regards the problem of learning the undirected graph underlying the MBV distribution which describes the presence/absence of 5 of the most common aeroallergens.
In particular, we analyzed a sample of 200 children between 3 and 12 years who had symptoms of inflammation of the upper and lower respiratory tract. The data was collected at the Department of Pediatric Allergology of the Policlinico Umberto I in Rome, Italy. For each child the positivity was measured for the most common aeroallergens (grasses, dust mites, olea, parietaria, alternaria) by means of prick tests evaluated after 15-20 min exposure with positive results defined as a wheal $\geq 3 mm$ diameter.
Figure \ref{fig:real_data} shows the results of applying our procedure to this aeroallergens data set. It is informative to examine the different graphs we obtained choosing different threshold in step 3 of our procedure. In particular, starting from the left upper graph, where no threshold is applied we end up to the bottom right graph where 75\% of the edge are killed by the threshold. The graphs in panels 1, 2, 3 and 4 of Figure~\ref{fig:real_data} are nested each other since the type of threshold chosen acts progressively zeroing more and more terms. Hence, from this type of analysis, researchers may decide as a first instance to analyze the molecular similarities between  olea (O) and parietaria (P) as well as between olea (O) and alternaria (A). 

This is just a real example of how the proposed method can be used, which obviously does not pretend to be completely resolutive but at least it allows  to get quite reliable solutions when the dimensional data regime is low.

\section* {Acknowledgments}
The author thanks Proff. Caterina Anania and Vincenza Di Marino of the  Pediatric Allergology Department of Policlinico Umberto I in Rome for having provided the data and valuable explanations in this regard.

This work was partially supported by Italian Flagship project InterOmics.


\end{document}